\documentclass[12pt]{book}
\usepackage[utf8]{inputenc}
\usepackage[T1]{fontenc}
\usepackage{csquotes}
\usepackage[margin=1in]{geometry}
\usepackage{array}
\usepackage[authordate,backend=biber, maxcitenames=1]{biblatex-chicago}
\begin{document}

\chapter*{Large Language Models, Machine-Assisted Cultural Analytics, and the Agentic Turn}

\vspace{-1em}

\author[\textbf{Andres Karjus} ]{ Tallinn University; Estonian Business School; University of Tartu}

\vspace{0.9em}

\noindent\textbf{Synopsis.} Technologies like large language models make it easier to convert cultural material into analyzable variables, while LLM-driven agentic systems extend machine assistance to other parts of the research workflow. Rather than a full how-to guide or an exhaustive review of recent LLM use cases, this chapter looks at some crucial aspects of this new paradigm as well as the bigger picture. I argue that these technologies expand the scale, methodological range, and scope for delegation, while the basic logic by which those measurements support cultural claims remains mostly unchanged.

\vspace{0.8em}

Cultural analytics turns forms and practices into analyzable data -- archives, databases and corpora of texts and other media, taxonomies, variables -- to investigate human socio-cultural behavior at scale.
Qualitative and mixed methods are also part of the field, but quantitative remains its most common mode. 
Some questions can be answered directly using countable manifest features such as words, lengths, or names, which may require only light parsing.
Other, often arguably more interesting ones, necessitate first operationalizing latent or interpretive content -- meaning, emotion, framing, visual motifs, etc.
Much of this involves classification in one form or another: sorting and annotating these various outcomes of human cumulative cultural evolution into predefined categories or taxonomies, so that patterns and relations can be systematically assessed \parencite{bamman_2024_classification,Karjus2025MAQD}.
This is often laborious: to convert or "quantitize" such material into comparable quantitative variables, a systematic coding scheme needs to be developed or adapted, and somebody needs to read or look or listen to each unit or example, and assign the predetermined labels or values -- that can then be quantified and modeled in turn. 
Debates around enumeration, scale and interpretation have also examined how methodological choices and biases shape the results that follow \parencite{Moretti2013DistantReading,Piper2019Enumerations,vannoord_2022_analytics}.
All this work may be done by the lead researcher themselves, coauthors, or delegated to research assistants or crowd-workers. For the majority of the history of humanities and social sciences, all of them were humans. 

Computational approaches in these fields have benefited from successive advances in computer science, particularly in machine learning (ML), natural language processing (NLP) and AI technologies, improving either digitization or analysis of ever more cultural and behavioral data. To name a few, corpus statistics and topic models enabled "distant reading", supervised ML classifiers the larger-scale analysis of content variables, visual object classifiers gave rise to big data multimodal studies, and word- or sentence-level language models enabled increasingly automated research into meaning and its change over time. Not everything has to be necessarily manually analyzed by humans any longer. Pretrained generative large language models are the next step in this line of useful computer science technologies that make flexible interpretive operations even easier to set up, scale, and repeat, broadening the scope in terms of viable sample representativeness and number of variables used to describe a construct. They are also highly unlikely to be the last step.

Automated data annotation and classification has been available for a while, primarily via supervised machine learning, either in the form of ready-made task-specific classifiers like sentiment analysis tools, or pretrained models further adapted to a task using human-annotated examples. When it comes to textual data, all these approaches belong to a longer lineage of language modeling and distributional semantics, often associated with the dictum "you shall know a word by the company it keeps" from \textcite{firth_1957_synopsis} -- the idea that aspects of meaning are revealed by recurrent patterns of linguistic context. Modern LLMs learn such patterns from very large training corpora in the pretraining phase. Earlier approaches continue to have applications, including n-grams, static word embeddings that represent lexical meaning in a vector space \parencite{pennington_2014_glove}, and contextual embedding models like BERT, whose representations are sensitive to surrounding text \parencite{devlin_2019_bert,ReimersGurevych2019SentenceBERT} and can support retrieval and classification via task-specific fine-tuning.

LLMs enable a particularly flexible form of "in-context learning": they can perform tasks from instructions alone ("zero-shot") or with examples in the prompt ("few-shot"). For many tasks this removes the need for task-specific training data altogether, or substantially reduces it when fine-tuning is still needed. This makes LLM use cases naturally closer to human-driven data labeling workflows that have already been used and refined for decades: the same annotation \textbf{instructions} or coding schemes that research assistants or crowd-workers would have received before can now be passed directly to an LLM (while the need for validation of course still remains). Comparisons across various research settings report roughly order-of-magnitude reductions in annotation time, with the realized savings depending on the task and subsequent review burden \parencite{long_2024_evaluating,jahan_2024_finding,liu_2025_voices}, and lower annotation costs at similar or even improved quality, depending on the human baseline \parencite{GilardiEtAl2023ChatGPTAnnotations,Karjus2025MAQD}. As with other computational approaches, the methodological question remains whether the resulting operation distinguishes the historical and cultural contexts relevant to the inquiry \parencite{underwood_2025_impact}. Many of the concepts discussed below therefore apply to other methods as well, but LLM usage brings some of them into sharper focus.

Using these often multilingual and multimodal models for data work is of course no longer a novelty in (the related and often overlapping) fields of computational or digital humanities, computational social science, and cultural analytics. The range of applications is broad and growing, encompassing research into e.g. literature \parencite{kurzynski_2026_cognitive} and folk tales \parencite{HobsonEtAl2024StoryMorals}, historical news topics \parencite{oiva_2024_framework}, documents and records \parencite{StewartSinha2025UnstructuredHistoricalSources,karch_2025_llm}, political manifestos \parencite{benoit_using_2026}, narrative networks \parencite{PiperXuRuths2024SocialLives}, media \parencite{parson_2026_language}, semantic variation and change \parencite{periti_2024_lexical,karjus_2024_evolving}, video semiotics \parencite{paekivi_2026_modeling}, religious iconography \parencite{SpinaciEtAl2025ChristianIconography}, visual communication \parencite{nemani_2026_large}, to name a few. Beyond benchmarking research tasks like annotation, there are of course entire literatures growing around questions of how LLMs "think" (artificial cognition), how they may be biased in various ways, and what LLMs can or cannot "do" compared to humans -- including in cultural activities of interest to humanities such as art or creative writing \parencite{begus_2024_experimental,bellemarepepin_2026_divergent,eisenmann_2026_expertise,koivisto_2023_best,doshi_2024_generative}.

It has been an interesting few years where one could observe conference programs and journal issues that occasionally consist almost entirely of titles essentially boiling down to "LLM can be used to do a thing with data", accompanied by a smaller parallel industry devoted to showing that sometimes it (or at least the current version) cannot. Mentioning one in the title is often no longer necessary, as it has become an (often reasonable) default for many such tasks. This is not entirely unlike the surges of popularity of topic models or embeddings in past decades (which all still have use cases too), but not quite the same either. LLM usage is arguably easier -- prompt instead of train -- and their applicability broader. The same model can be used to filter a dataset, normalize it, and annotate it for dozens if not hundreds of variables. But the same model may also be powering a tool that will code the scripts to do these things, and write up the results.

One metascientific implication going forward is that dataset size and construct complexity as such will likely lose importance as selling points, while the driving idea, theory and method building, bias and uncertainty assessment, replicability, and real-world impact become the more consequential criteria for assessing the value or contribution of a given study. 
This shift likely also has implications for the identity of "digital" humanities and culture scholars: computer-science adjacent technical skills like basic programming that once constituted a marker of a certain scholarly profile may become less differentiating as such capabilities become broadly available through AI and agentic systems.

The focus of this chapter was originally intended to be on LLMs and their outputs as instruments for data quantitization, and to discuss related technical concepts. However, one cannot look past the fact that using LLMs for annotation is merely one aspect of a broader emerging paradigm of \textbf{machine-assisted research}, in which tasks are increasingly described and delegated not only to human assistants or colleagues, but also to artificial agents. Yet, cultural analytics and the aforementioned adjacent interdisciplinary fields, with their history of near-constant flux, should be fairly well prepared for the changing research landscape ahead.

\section{LLM-Assisted Quantitization, Annotation and Classification}

The term "large language model" is usefully narrower than "AI", which names a field, but no longer quite captures either the increasingly multimodal capacities of these models or the larger systems built around them. It is therefore useful to distinguish the model itself from the interface and workflow in which it operates. The model produces representations or outputs, ranging from text and annotations to tool calls activating other models or software. The interface is how a researcher accesses it, through a chat window, integrated development environment (IDE), or application programming interface (API). A pipeline may combine model calls with operations such as preprocessing, retrieval, and statistical analysis. An agentic system goes further by allowing model outputs to determine subsequent actions in response to intermediate results, including delegating subtasks to other agents.

This section focuses on perhaps one of the most common current use cases of LLMs in cultural data analytics, the systematic \textbf{quantitization} or conversion of qualitative data (such as text) into predefined units and fixed categorical or numerical variables for subsequent quantitative analysis (as exemplified by the cited works above). Related practices have long existed in content analysis, mixed-methods research, and feature analysis in corpus and functional linguistics. 
Terms such as annotation and coding describe related operations (but do not necessarily imply subsequent quantification). In a quantitizing design, the research question is answered primarily through quantitative analysis of the resulting variables rather than annotation or inspection of labels alone. The activity of looking at a unit of data such as a sentence, and determining the label(s) to assign to it can range from a quick binary decision to an entire in-depth qualitative analysis of that unit, to filling in values for hundreds of categorical variables. In that sense this is mixed methods: qualitative analysis of units paired with the quantitative analysis of the derived variables.
Notably, this approach is the default in many domains to the extent that it is often not named as a specific method. 

A typical workflow proceeds from a research question or hypothesis and source data collection through unitization and codebook development to annotation, statistical analysis, and finally, interpretation. Importantly, the statistical claims are based on quantitative modeling of the quantitized variables, while qualitative inspection of the raw data remains useful for interpreting those results and illustrating them with examples. These stages are often iterative, with validation accompanying codebook development and annotation rather than being reserved for the end. To avoid overfitting or accidental data dredging, data used to develop the instrument should be kept separate from those used for confirmatory analysis.

LLMs slot into this quite nicely, particularly in the annotation phase, doing the work that otherwise would be left to humans or purpose-built classifiers -- with the advantage that they can often receive essentially the same instructions as humans, but unlike conventional classifiers, may return malformed or unexpected answers. Structured outputs and similar validation checks, however, increasingly help control these issues (but note that this is a different failure mode from classification error or human-machine disagreement).

\subsection{Sources, Units, and Coding Schemes}

Before deciding what to delegate to an LLM, the research question needs to be translated into a usable coding or variable scheme, and data into comparable units. 
Unitization connects selected sources to the intended inference. For some questions, the unit is obvious. If words are to be studied, the natural unit is a word (perhaps in context). Sentences or paragraphs often make sense when studying text and are easy to segment. A painting is often a natural unit when studying art, though artist is the likely grouping that probably requires suitable statistical modeling down the line. 
But a stance may emerge across several sentences, a character relation across a scene, or an iconographic motif within one image region. 
Various aspects like corpus composition and exclusion logic, optical character recognition (OCR) quality, transcription and metadata reconciliation matter for unitization design \parencite{rawson_2019_against}. Inherited metadata and institutional labels embody earlier decisions and should be considered, e.g. when using an archive's identity categories \parencite{gitelman_2013_raw}.

Earlier language models were constrained in the amount of context they were capable of handling as one unit, but longer context windows, by now in the millions of tokens, widen these options. Too large a unit containing extraneous material can confuse or mislead both human and machine annotators, however. One additional solution is to create an intermediate targeted representation of a larger document, a structured abstraction or a summary, before the annotation step \parencite{benoit_using_2026}.

The annotation or coding scheme or codebook -- note that naming this entity varies across disciplines -- specifies which features or variables are to be annotated for each unit, and the allowed values of each variable. Readily countable properties remain useful where they answer the question \parencite{Moretti2013DistantReading,Piper2019Enumerations}; more abstract constructs require an account of what would support or contradict a proposed reading \parencite{radford_2020_theory,strauss_2009_construct}. 
For e.g. \textit{identity}, relevant dimensions might include institutional references, narrative roles, evaluative terms, or recurrent contrasts. Making the concept empirically tractable does not exhaust its meaning, and persistent mismatches may call for revising the theoretical category rather than (only) adjusting the annotation instrument or LLM prompt \parencite{algeehewitt_2024_computing}.
The coding scheme should be built with down the line statistical modeling in mind, with an understanding of the role of each variable (predictors, controls, the dependent variable of interest).
A common \textbf{pitfall} in quantitizing (and also survey) research is to collect many variables with numerous categories or response levels, and then afterwards try to devise a statistical test for the research question, leaving more parameters and analytic degrees of freedom than the sample can reliably support.

Clear instructions make coding distinctions usable by both human and machine annotators, through definitions, examples, boundary cases, allowed evidence, and ways to record uncertainty \parencite{piper_2026_theoryfirst}. Some constructs may be represented adequately by a single variable, while others require several dimensions. A coding scheme should make explicit what each variable captures and what it does not, so that different aspects are not collapsed into one label nor counted twice under different names.
E.g. a book review may evaluate a novel positively in aesthetic terms but negatively in political terms; if both matter to the research question, they are better represented as separate variables than collapsed into one overall positive-negative judgment. Same goes for levels of a categorical variable: that binary contrast might be too crude, while a dozen might be cumbersome to model down the line, and also difficult for the annotator to distinguish.

\subsection{ Machine-assisted Annotation at Scale}

A major advantage of LLM-assisted annotation is that it makes coding at corpus scale feasible for many tasks that would otherwise be prohibitively time-consuming or costly. But large collections need not always be annotated exhaustively with the largest most capable LLMs. Genres or materials outside the research scope can be excluded, while keyword or embedding matching or a small, fast classifier can identify promising candidates for annotation. To retain collection-wide estimates, a probability sample from the remaining material can preserve coverage while concentrating annotation effort where it is most useful \parencite{imberg2025active}.
One option is to sample within retrieval-score strata, oversampling promising units (e.g. by keyword or embedding similarity) while retaining known, nonzero inclusion probabilities for every unit. Recording these probabilities and stratum sizes allows subsequent weighting, although very small samples from large low-scoring strata may still leave estimates imprecise (which should be taken into account in the statistical modeling side). Retrieval then allocates effort without defining the target category's presence.

Note that this discussion pertains to LLMs and pipelines that enable direct control of the model and its parameters, using it as any other machine learning model. Chatbot interface, while popular, are less suited for such tasks, as they are optimized for, well, chatting, often set up with high temperature (output randomness) and to take previous messages in an interaction (chat) into account when generating the next output, and come with opaque system prompts in the commercial cases. While a systematic annotation process requires control over these aspects, fixed output format, replicability, and in the vast majority of cases, independence between annotation passes and past decisions.

In the annotation loop, each selected unit (with or without additional contextual content), and the coding instructions are sent to a model as input (or several models in an e.g. majority voting design). This could be an LLM running either locally, in a compute cluster, or via a dedicated LLM cloud service API (most large LLM production companies also offer paid APIs). The returned outputs are linked back to their sources. Model choice follows the task: instruction-tuned LLMs can apply a scheme without task-specific training. Their performance on a specific task cannot always be assumed from general benchmark-performance \parencite{bamman_2024_classification,ZiemsEtAl2024CSS}, so creating a focused expert-annotated test set is often reasonable. But smaller supervised or fine-tuned models also can be competitive even with few labeled examples, especially on simpler and repeated tasks \parencite{TunstallEtAl2022SetFit,AlizadehEtAl2023OpenSourceLLMsAnnotation}. Setups vary and should first and foremost follow the research question and coding scheme.
Examples include direct coding of e.g. thematic and linguistic text or visual features \parencite{Karjus2025MAQD,SpinaciEtAl2025ChristianIconography}, metadata extraction and entity resolution \parencite{StewartSinha2025UnstructuredHistoricalSources,BisianiEtAl2025GeoparsingUKLocalMedia}, but may also draw on intermediate representations, such as image descriptions \parencite{Thomas2025AutomatingIconclass} or issue-focused summaries \parencite{benoit_using_2026}; LLM-generated labels can also be combined with topics and entity mentions to model co-occurrence \parencite{dobranic_2026_approaches}.

\section{Annotation Reliability and Modeling Uncertainty}

When it comes to humanities and cultural subject matter, quantitization is often at least partly interpretive or subjective. Human annotators in these domains therefore rarely agree perfectly, and there is a long tradition of handling this variation \parencite{krippendorff_2004_reliability}. LLM-assisted annotation extends many of the same questions to machine annotators: where do judgments diverge, by how much, and does that variation matter for the eventual comparison? A human consensus label is itself usually an aggregation of judgments rather than direct access to ground truth. Reliability is therefore relative to the intended use of the annotated variables from the quantitization procedure: evidence may be sufficient for a bounded corpus or task, while comparisons across periods, languages, genres, etc. entail checking whether the analysis remains comparable across those settings.

\subsection{Agreement, Benchmarking, and Error}

Human reference coding or benchmarking is a useful source of evidence, but need not always mean exhaustively validating every level of every variable, where resources are constrained. A smaller test set can be used to assess whether machine annotations are sufficiently close to human expert ones \parencite{CalderonEtAl2025AltTest}. Model-to-model agreement can likewise help identify tasks or cases for which machine annotation is relatively stable \parencite{AhmedEtAl2025ManualAnnotation}. Where several related variables are coded under the same scheme, evidence from a subset may reasonably inform the rest if their difficulty and error structure appear comparable, although this adds uncertainty and is best treated as a calculated tradeoff.
Another option is to treat several LLMs, prompt variants, and/or human annotators as a panel of raters rather than designate one source as truth. Their judgments can be aggregated or retained separately, with disagreement itself providing information about reliability \parencite{than2025qualitative,AhmedEtAl2025ManualAnnotation}. Where disagreement reflects meaningful differences in interpretation rather than simple error, retaining that variation may be more informative than collapsing it into a single majority label \parencite{piper_2026_theoryfirst}.

Multi-model (vis-à-vis multi-human) agreement can be useful evidence, but does not necessarily imply independence. Different LLMs can make correlated errors even across providers and architectures \parencite{kim_2025_correlated}; agreement may reflect e.g. shared training data or cultural biases. What counts as relevant agreement also depends on the intended interpretation: an LLM may be asked to approximate a domain expert, a defined group of annotators, or a distribution of perspectives, and these are not interchangeable targets. Validation is therefore most informative when the reference annotators correspond to the kind of judgment the resulting labels are meant to represent. Using models from different families, varying prompts, and occasional targeted human checks can reduce this risk without requiring a large gold-standard set for every variable. In the end, it is a \textbf{balancing act} between time sensitivity, human annotation budget, construct complexity or ambiguity, and how sensitive the final result is to plausible alternative labels. A study without external validation of its LLM annotations is not inherently invalid. Rather, the validity of its annotations and robustness of its conclusions are less well established, including whether the coding would hold across annotators, models, or samples; this is not entirely unlike a dataset annotated by a lone human expert (as is common in many disciplines).

Annotation uncertainty can and usually should also be carried into the statistical analysis. Where several raters are available and a shared underlying category is assumed, latent-rater models such as Dawid--Skene can estimate coder-specific error rates and probabilities for the underlying labels \parencite{dawid1979observer}; multiple imputation can then propagate uncertainty in those labels and rater parameters into downstream estimates \parencite{boeschoten2017milc,boeschoten2018updating}. In that model, ratings are conditionally independent given the underlying category and rater-specific error parameters. Other approaches use a smaller reference sample directly. Design-based Supervised Learning (DSL) combines machine predictions with a probability sample of reference annotations to correct downstream estimators \parencite{EgamiEtAl2023DSL}. Confidence-Driven Inference (CDI) uses confidence information and observed errors to guide human sampling while accounting for the selection design \parencite{gligoric2025unconfident}. These corrections operate relative to the reference annotations, so the construct validity of that reference still requires justification. Simpler bootstrapping or simulation of an estimated confusion structure can be an easier fallback, although it relies more directly on the assumption that the estimated error process carries over to the full dataset. But in the end, high agreement or classification accuracy does not itself guarantee valid downstream inference alone \parencite{baumann2025llmhacking}.

\subsection{Robustness Across Contexts}

An aggregate evaluation score answers only part of the question of how an annotation procedure behaves. A reference sample drawn from the material of interest can estimate its average performance on that material, but aggregate performance may conceal concentrated or systematic failures. Targeted behavioral tests can be used to check whether the annotator responds sensibly to controlled changes in the input, e.g. whether a label remains stable after an irrelevant wording change or changes as expected when relevant evidence is altered \parencite{ribeiro_2020_beyond}. Such challenge cases are diagnostic rather than estimates of corpus-wide error. A high failure rate on a deliberately hard test identifies a weakness, but not how frequently it occurs in the collection.

Comparability across languages, genres, periods, or communities depends both on whether the category refers to the same phenomenon and on whether the model handles the different material similarly. Cultural categories may rely on different cues or meanings in different settings \parencite{kommers_2026_computational,ZhouBammanBleaman2025CultureNotTrivia}. Model behavior can also shift with input, and even prompting or fine-tuning on period prose might not necessarily produce historically well-aligned outputs \parencite{underwood_2025_can}. In comparative studies, it might be worth checking model performance separately for the groups or strata being compared. Where the construct itself differs across them, the coding scheme may also need adaptation, provided the resulting variables remain comparable.

Once the intended construct and scope are defined, robustness can be further assessed across defensible implementations of the same instrument. Coding scheme and subsequently, prompt wording, label order, few-shot examples, or model differences may alter individual annotations -- the question is whether results and conclusions change across theoretically defensible variants \parencite{piper_2026_theoryfirst}. Variation within the variants and prompts can also be informative.
The same evaluation material can remain useful as models and implementations change. Cases that expose known failure modes can be retained in a versioned test set. Shared resources like benchmarks and shared tasks \parencite{schlechtweg_2020_semeval} can support the same cumulative work across projects, as is common in more computer science adjacent fields.

This all is not to say one should always exhaust every possible validation or robustness check. A reasonable stopping point is usually when the evidence is proportionate to the breadth and depth of the intended claim, and any remaining uncertainties and limitations can be transparently reported.

\subsection{From Model Outputs to Evidence}

Once quantitization has produced categorical or numerical variables, they enter the same kinds of quantitative analyses as variables obtained through human coding or other measurement procedures; little is specific to LLMs at this stage. Descriptive claims can be limited to the observed corpus; claims about a broader population require an account of how the analyzed material relates to that population, usually together with a statistical model that estimates patterns and their uncertainty under explicit assumptions. Such a model can represent dependence and quantify uncertainty, but it cannot by itself resolve source-coverage or selection bias. The model should reflect the structure of the observations, including relevant dependence, e.g. passages clustered within authors or periodicals, which can easily be modeled using a mixed-effects or hierarchical model \parencite{grimmer_2013_text,arnold_2019_new}. LLM-assisted coding may be seen as adding another source of measurement uncertainty, to be considered alongside sampling and statistical uncertainty.
For example, a corpus annotated for genre or style might suggest that one category becomes more common over time. Interpreting that pattern still depends on what was sampled, what counted as the category, whether the coding behaved similarly across periods, and what denominator the estimate refers to. Even a robust trend still describes a pattern in the \textit{measured data}; it does not by itself identify the cultural process that produced it.

LLMs can also support qualitative work by summarizing interviews, proposing themes, comparing passages, or helping researchers move between cases and candidate concepts, functioning as \enquote{interpretive screens} \parencite{ringler_2024_computation}. Comparison need not be numerical, and summaries need not become variables to be useful. Quantitative results can in turn guide the selection of material for close reading \parencite{pianzola_2020_wattpad}.
The weaker \enquote{quasi-quantifying} practice arises when counts or impressions of prevalence in a limited sample are treated as broader substantive findings without quantifying the relevant sampling, coding, or model uncertainty. Close examination of sources can of course help explain the resulting findings, annotation errors, and unusual cases, while quantitative analysis should indicate how strongly the result is supported across the material and how much variation might be expected if the analysis were repeated on comparable data.

\section{Agentic Systems and the Redistribution of Expertise}

Agentic AI systems or harnesses are understood here as software that can autonomously execute parts of a workflow, without direct human direction.
Examples include calling external tools, models, or subagents, iterating over intermediate results, waiting for external results and resuming execution, and maintaining a persistent control loop. The underlying LLM may still be an autoregressive input--output model which repeatedly predicts the next token from the available context until generation stops. 
The agentic shift is from such fixed-control-flow inference or pipelines toward adaptive control flow, where model outputs can determine what operation happens next, until some goal is achieved.
Results from search operations, code execution, data transformations, or calls to other models can inform subsequent actions and reasoning \parencite{yao_2023_react}. These setups can therefore extend machine delegation to more complex, multi-step research tasks: proposing units or codebooks, retrieving and synthesizing literature, writing and checking code, producing experimental interfaces, analyses and figures, and reviewing intermediate results \parencite{ciotti_2026_agenda,karch_2025_llm,asai_2024_openscholar,liang_2023_can,boiko_2023_autonomous}.

Current software examples driving this agentic turn include Codex by OpenAI, Claude Code by Anthropic, and open-source alternatives like Hermes Agent and OpenClaw.
This is qualitatively different from simple turn-based chatbot use, where a model is invoked once per user turn, as well as typical annotation work based on independent model calls. However, data annotation is likely still best implemented as per-unit calls, so that each unit of data is analyzed on its own without influencing later ones; an agentic system can easily set up such loops too.

The division of labor between humans and machines may vary. The scope can extend from a research idea or human-guided data work to complete research pipelines culminating in complete manuscripts drafts \parencite{lu_2024_ai,ifargan_2025_autonomous,engzell_2026_paper}. Producing an analysis and reporting it in the form of a paper or book does not by itself establish scholarly quality, and in a recent evaluation of research reproducibility tasks, human researchers outperformed the tested minimally supervised agentic workflow \parencite{brodeur_2026_ai}. At least for now. So the question becomes, what to delegate. While public benchmarks give some intuition about general capacities of LLMs and the systems running on them, identifying useful applications for one's own research entails some trial and error.

Now that quantity is cheap, quality, not just speed and volume, should increasingly become the focus of evaluation and (both internal and peer) review \parencite{ma_2024_dancing}. Polished prose can conceal errors and missteps; making code, data, logs, and decision paths useful for assessing results \parencite{zhang_2026_how}. In addition to the typical code and data attachments accompanying published papers, information about the AI systems used should likely also be reported. And obviously the human researcher still retains responsibility in the end; it is unlikely that "my AI did it" would ever become a valid excuse for irreproducible or botched research.

Agents also make complex implementation and quantitative designs accessible to researchers with little to no such experience. Running an analysis one does not fully understand can of course lead to mistakes or entirely spurious results, but exhaustive implementation knowledge has not been entirely necessary for computational research either. Rhetorically: how many people reading this chapter can honestly say they understand the minute mathematical details of a regression optimizer, the architecture of the neural network classifying their text, or indeed have manually read and thought deeply about every single text or image that they have processed using computational tools? Perhaps some. But the relevant assumptions and interpretation should still be understood well enough to judge what the results support.

Besides gains in work efficiency and access to powerful data annotation tools, there is another interesting, perhaps sociological or meta-scientific angle here. The ability to carry out certain analytic tasks has helped delineate the scholarly \textbf{identity} of fields like digital humanities, computational social science, and cultural analytics. In the recent past, an average qualitative humanities or cultural studies researcher likely could not easily wrangle large datasets, train a custom classifier on an HPC cluster, or interpret complex quantitative outputs on their own; people who could do so often attached words like digital or computational to their job descriptions, conference titles and journal names, while still dealing with largely the same subject matter and questions about human culture and behavior. Tasks outside one's expertise could be delegated to capable students or coauthors, but not everybody had that luxury, and the scarcity of technical skills carried its own academic value.

AI now changes that division of labor. A single researcher or small team can increasingly combine forms of technical and disciplinary work that would previously have required several people with complementary expertise. AI assistants can also support interdisciplinary collaboration by helping to translate differing theories, terminologies and approaches between colleagues who do not share the same background. This broadens access to ambitious interdisciplinary research and changes what counts as a feasible project for a given team size -- perhaps faster than institutional expectations around staffing and feasibility can adjust.

For the price of a dinner or a movie ticket, anybody with a computer can subscribe to agentic tools that write thousands of lines of code per minute, use top-of-the-line artificial intelligence in the cloud, summarize and synthesize thousands of prior works, design sophisticated statistical and machine learning pipelines, and write up the results in perfect (if a tad repetitive) English or any other language of choice. Some of it won't be perfect. But neither is the history of the humanities and social sciences (see: replication crisis). The technical complexity of a study may no longer necessarily remain a hallmark of what editors and reviewers will consider sufficiently substantial or novel, nor of the expertise of the author. This is also not necessarily a bad thing as such. Research questions involving big data and methods capable of handling many interacting variables become accessible to a wider range of researchers -- useful when our subject matter itself is often complex, because human culture, society and language are inherently complex systems \parencite{fivegraces_language_2009,miller_complex_2007}. While science rewards ambition, risk, and quality over quantity, academia... is a different beast. Widespread AI access may well support the increase of easy, safe, incremental, yet marginally publishable studies \parencite{metcalf_2026_aiproduced}.

As an anecdotal example from personal experience: when conducting occasional AI workshops or courses for academic audiences, I now routinely run the live demo of asking for an example research question, and then getting an agentic system like Codex to generate a complete manuscript draft from scratch while I continue with the lecture or practical exercises. There, agents quietly design and coordinate between themselves tasks like research design, data collection, wrangling and annotation as needed, statistical modeling, figure production, manuscript write-up, suitable venue lookup and selection, journal template application, cover letter preparation, the compilation of a synthetic review panel based on public background information and likely pet peeves of the editors and researchers likely to review for the selected journal, followed by a few iterations of internal pre-review and improvements, setting up a scheduled task monitoring my inbox for a rejection or resubmission response, and uploading the manuscript to the journal submission system, save for clicking the final "submit" button. It usually takes under an hour and requires little to no human intervention. Would a highly selective venue publish it? Of course not. A low-impact one? If not the first, then eventually one will, and the resubmission process can easily be automated, too. As reviewers, we are regularly already seeing the other end of this process, as our inboxes overflow with weekly if not daily (increasingly automated looking) review requests.

This example may be extreme -- only one step shy of replacing the aforementioned audience input as well with an agent that continually generates research questions and delegates them to subagents for milling \parencite{engzell_2026_paper}.
At the same time, reasonable and controlled agentic AI is clearly useful, \\
automating mundane tasks (who can say they enjoy the data entry on each new journal submission or project report?), \\
speeding up data annotation (once a quantitization scheme and an annotator model have been validated), \\
leveling the academic writing playing field for non-native English speakers,\\
and enabling access to complex processing and modeling to those without a second degree in computer science. 

The initial draft of this chapter was generated by Codex running GPT-5.5, followed by substantial human revision, where the assistant (now on GPT-6) acted as more of an interactive thesaurus and spellchecker. A fully agent-written handbook is nevertheless easy to imagine, and likely many already circulate. The more interesting shift is therefore not whether machines can produce complete scholarly artifacts, but how this changes the division of research labor and the expertise worth developing and retaining.
To stay competitive in academia, the cultural scholar will likely need to eventually complement artisanal coding and analytics with some degree of automation. But in some sense, agentic AI is also just another computational tool in a long line of such tools. While some of our humanities colleagues continued to engage in close reading, we increasingly had machines handle digitization, the counting of words in corpora, then, modeling meaning. From this perspective, extending the same logic to further parts of the research process is more a continuity than rupture.

Another aspect besides disciplinary identity concerns the role of hands-on practice and craft in scholarly expertise. For example, as another personal example, I quite enjoy finding and wrangling datasets, deciding how to analyze them, fiddling with figures in ggplot2, and reading interesting related research literature. 
Delegating some of these can free time for interpretation and perhaps thinking about more consequential questions, while also feeling like a loss of part of the craft. At the same time, doing the work directly can itself develop the judgment needed to evaluate similar work later.

Such delegation -- even of simple data annotation or data entry -- may also displace paid work and remove opportunities through which students and junior staff learn the sources, research designs and analysis practices. Delegation risks deskilling both novice and senior scholars and may further erode the capacity to evaluate the work being accepted. Evaluation requires enough programming knowledge to check implementation and data operations, enough statistics to assess an inference and its assumptions, and enough domain knowledge and source criticism to reject plausible but poorly grounded results. That expertise need not reside in one person, but should ideally remain available within the research team.

But the role of a manager is nothing new in science either -- it just used to be the prerogative of principal investigators: specifying work, delegating it, and remaining responsible for the result. What changes is partly who, or what, performs the delegated work. The teaching question follows: how much limited curriculum time should go to, for example, programming, and how much to delegation, directing and evaluating systems that can program? Cultural analytics has experience adapting computational methods to humanistic questions; that history hopefully offers a basis for deciding critically which activities to delegate and which remain valuable to perform and teach directly.

\section{Infrastructure and Research Design}

LLMs and systems that use them are relatively computational-resource hungry. They can be run locally, on institutional or private HPC clusters, or through dedicated API services. Hosted commercial frontier models make capabilities available without the substantial hardware and maintenance required to run them locally, but leave access, model continuity, and pricing largely with the provider \parencite{burkhardt_2024_foundation}. API-only models may also be updated or discontinued, sometimes changing the performance of an otherwise unchanged workflow \parencite{ma_2024_why}. Local or open-weight models give researchers more control over the working configuration, but shift more of the computational and maintenance burden back to the research side. Which arrangement makes sense therefore depends on available resources, legal constraints on the data, and the role of LLMs in the pipeline.

That role also determines how much exact model continuity matters. If an LLM is used as a measurement instrument to produce a large annotation set, retaining the model version, prompts, outputs, and evaluation material has obvious value for both internal consistency and continuity as well as reproducibility. If it serves mainly as an exploratory or coding assistant, exact replay of every interaction may be less important than preserving the resulting data, code, and analytic decisions. Human-only research does not normally reproduce the exact judgments of its original annotators either; e.g. when replicating a study, it would be unreasonable to require the exact same group of research assistants, crowd-workers or the original author to re-code the task. A more useful standard for both approaches is whether the procedure is documented, and its uncertainty understood well enough for the design to be replicable, and the substantive result auditable. 

\enquote{Open} is however a scale, both in science and model training. Code, model weights, training and alignment data, licenses, documentation, and access may each be available to different degrees \parencite{liesenfeld_2023_opening,widder_2024_why,klein_2025_provocations}. Access to weights (the substantive component of an LLM) can make a model easier to preserve and inspect without revealing the data and decisions that produced it. For cultural analytics, those data are relevant because LLMs reorganize patterns learned from large collections of human cultural production \parencite{farrell_2025_large}. Training corpus construction and filtering can  change what is represented in such data \parencite{dodge_2021_documenting}, while this information on the most capable models typically remains only partly documented or inferrable at best. Some inherited biases or absences can be tested for, while others may only become visible in particular applications -- an example of an \textbf{unknown unknowns} problem. And as machine-generated material increasingly mixes with human-produced culture and later training data, the origins of particular content and patterns may become harder to trace \parencite{brinkmann_2023_machine}.
This is already relevant in studying Internet materials and crowdsourced surveys, where texts and images may themselves have been generated or altered with generative models, complicating what exactly is being measured \parencite{rilla_2026_pollution}.

There is also an interesting field-level consequence to relying on a relatively small, centralized set of models. While the variety of LLMs is large and ever-growing, the number of truly capable models -- fit for the most demanding analytics tasks -- is significantly smaller, and the number of their producers even more so, reflected by concentrated usage observed in recent meta-scientific surveys  \parencite{rossi_2024_problems,guan_2026_generativeturn}. 
While reliance on on a few popular architectures (like LDA for topics) or pretrained models (like BERT) is not new in DH and cultural analytics, there may be even more concentration here. This is somewhat understandable: while an older approach or a small LLM may well suffice for simpler annotation or tagging tasks, frontier models that can be called via APIs using a couple of lines of code are a rather obvious easy choice for solving challenging questions at big data scale, as their per-token usage costs have also gone down year by year.
Otherwise independent projects can therefore share a significant part of their measurement machinery -- harmless at best, but at worst a source of compounding biases and errors \parencite{kim_2025_correlated}. 
More broadly, widespread use of largely the same AI infrastructure may encourage convergence in analytic workflows and in the questions that are easiest to operationalize \parencite{MesseriCrockett2024Illusions,traberg_2026_ai}. 
It can also introduce a peculiar \textbf{circularity}. E.g. in the online culture studies example, researchers may use an LLM to analyze material assumed to be human-produced, when it may in fact have been generated by the very same popular model.
Using the strongest available model for a given task may nevertheless be entirely rational for an individual project; the concentration is likely to become visible across many studies than within any one of them.

The practical response is familiar \textbf{open science}: recording the data and sampling procedure, codebook and prompts, model and its producer (and provider, if different), version or date of access, relevant parameters, validation results, retained inputs and outputs where feasible, and analysis code \parencite{tornberg_2024_best,liesenfeld_2023_opening}. Copyright, privacy, platform terms, community obligations, or the absence of a suitable open model may limit how fully a workflow can be shared or reproduced. Full openness is not always feasible, and its absence does not by itself invalidate the research. The pragmatic aim should be to aim to do one's best: make the work as transparent and reproducible as the project and data reasonably permits, while stating clearly what could not be shared and why.

\section{In Summary}

Cultural analytics has repeatedly expanded its toolkit by making use of new computational tools. LLMs are the latest step in that history, but in their flexibility significantly broaden what can be delegated to a single general-purpose model. Like earlier machine-learning tools, they enable processing more material at scale; unlike earlier task-specific classifiers, they can often be used to apply multidimensional coding schemes directly from instructions similar to those given to human annotators. This makes a wider range of theoretical constructs tractable across larger samples and cultural data collections.

The basic logic of research, however, remains the same. What changes is where more of the work happens. While (ideally validated) machines take over routine quantitization tasks, more effort can move from carrying out every annotation decision by hand toward deciding what should be measured, translating that into well delineated variables, and thinking about what it all means and what we wish to learn about the world, culture, and societies.
Human and LLM annotators are not equivalent, but they can often occupy similar roles in this process. But neither needs to nor should be treated as an oracle: 
the relevant question is whether a self-conducted or delegated annotation work is reliable enough for the claim being made, and whether the remaining uncertainty is properly reflected in the subsequent analysis and interpretation.
In some cases, variance may also be meaningful itself.
But in the end, validation is a matter of proportion rather than perfection: there is rarely one ideal amount, only a reasonable amount given the resources available and the question being asked.

LLM-driven agentic systems further extend this delegation potential beyond annotation to other parts of the research project. As implementation becomes easier to hand off, technical expertise still matters, but more so in recognizing consequential questions, gaps, and connections, turning them into good research designs, and validating the resulting evidence. Technical, statistical, and domain knowledge still matter because they shape what can be imagined as well as what can be trusted. Models and interfaces will keep changing, and their capabilities will almost certainly expand. 
For cultural analytics, this expands the empirical reach of ambitious theory, empowering research at a scale and level of detail that can match the complexity of cultural phenomena.

\printbibliography

\end{document}